\documentclass{sf2a-conf}
\usepackage{graphicx}
\usepackage{bm}

\def\ba{{\bm a}}
\def\bb{{\bm b}}

\def\lb{\label}

\def\be{\begin{equation}}
\def\ee{\end{equation}}
\def\bea{\begin{eqnarray}}
\def\eea{\end{eqnarray}}

\begin{document}

\TitreGlobal{SF2A 2006}

\title{Relativistic astrometry and Synge's world function}
\author{Teyssandier, P.}\address{SYRTE, UMR/CNRS 8630, Observatoire de Paris, France.}
\author{Le Poncin-Lafitte, C.}\address{Lohrmann Observatory, Dresden Technical University, Dresden, Germany.}
%
\runningtitle{Relativistic astrometry and Synge's world function}
\setcounter{page}{237}
\index{Teyssandier, p.}
\index{Le Poncin-Lafitte, C.}


\maketitle
\begin{abstract} 
Almost all of the studies devoted to relativistic astrometry are based on the integration of the null geodesic differential equations. However, the gravitational deflection of light rays can be calculated by a different method, based on the determination of the bifunction giving half the squared geodesic distance between two arbitrary points-events, the so-called Synge's world function. We give a brief review of the main results obtained by this method.
\end{abstract}
%
\section{Introduction}

With advances in technology, it will become indispensable to determine numerous relativistic effects 
in the propagation of light beyond the first order in the Newtonian gravitational constant $G$, in particular 
in the area of space astrometry. The Global Astrometric Interferometer for Astrophysics (GAIA, Perryman {\it et al.} 2001) is already planned to measure the positions and/or the parallaxes of celestial objects with typical uncertainties in the range 1-10 $\mu$arcsecond ($\mu$as) whereas the Laser Astrometric Test Of Relativity (LATOR) mission will measure the bending of light near the Sun to an accuracy of 0.02 $\mu$as (Turyshev {\it et al.} 2004). In this last case, it is clear that the effects of the second order in $G$ must be taken into account. To obtain a modelling of the above-mentioned projects, it is necesssary to determine the deflection of light rays between two points $x_A$ and $x_B$ of space-time.  
In almost all of the theoretical studies devoted to this problem, the properties of light rays are determined by integrating the differential equations of the null geodesics. This procedure is workable as long as one contents oneself with analyzing the effects of first order in $G$, as it is proven by the generality of the results obtained in the litterature (Klioner 1991, Klioner \& Kopeikin 1992, Kopeikin 1997, Kopeikin \& Sch\"{a}fer 1999, Kopeikin \& Mashhoon 2002, Klioner 2003). Unfortunately, analytical solution of the geodesic equations requires cumbersome calculations when terms of second order in $G$ are taken into account, even in the case of a static, spherically symmetric space-time (Richter \& Matzner 1982, 1983). However, an alternative approach exists and seems to be promising. Based on the Synge's world function and variational properties of geodesic, it precisely does not require the knowledge of the geodesic and directly provides the time delay of light and the direction of a ray at the point of reception, i.e. at the observation point. In this work we derive the general expression of the angular separation between two point light sources as measured by a given observer in arbitrary motion. We show that the angular distance is fully determined if we calculate several ratios which can be obtained from the knowledge of the Synge's world function.

Throughout this paper, $c$ is the speed of light in a vacuum and $G$ is the Newtonian gravitational constant. The Lorentzian metric of space-time $V_4$ is denoted by $g$. We adopt the signature $(+ - - -)$. We suppose that space-time is covered by some global coordinate system $x^{\alpha} = (x^0, x^i)$. We assume that the curves of equations $x^i =$ const are timelike in the neighbourhood of the observer. This condition means that $g_{00}> 0$ in the vicinity of the observer. We employ the vector notation $\ba$ in order to denote either the ordered set $(a^1, a^2 , a^3)$, or the orderer set $(a_1, a_2 , a_3)$. Given  $\ba = (a^1, a^2 , a^3)$, for instance, $\ba . \bb$ denotes $a^i b^i$ if $\bb = (b^1, b^2 , b^3)$ and $a^i b_i$ if $\bb = (b_1, b_2 , b_3)$), the Einstein convention of summation on repeated indices being used in each case. The quantity $\vert \ba \vert$ denotes the ordinary Euclidean norm of $\ba$ : $\vert \ba \vert = (\delta_{ij}a^i a^j)^{1/2}$ if $\ba = (a^1, a^2 , a^3)$, and $\vert \ba \vert = (\delta^{ij}a_i a_j)^{1/2}$ if $\ba = (a_1, a_2 , a_3)$. The indices in parentheses characterize the order of a term in a perturbative expansion. Theses indices are set up or down, depending on the convenience.

\section{Angular distance as measured  by an observer in arbitrary motion}

To begin with, let us consider a light ray $\Gamma$ received at point $x_{o}=(ct_{o}, {\bm x}_{o})$ and let us recall how is defined the direction of this ray as measured by an observer ${\cal O}(u)$ moving at $x_{o}$ with a unit 4-velocity $u$. The three-space relative to the observer ${\cal O}(u)$ at point $x_{o}$ is the subspace $\Pi_{x_{o}}^{(3)}(u)$ of tangent vectors orthogonal to $u$ (see figure below). 
\begin{figure}[h]
   \centering
   \includegraphics[width=13.5cm]{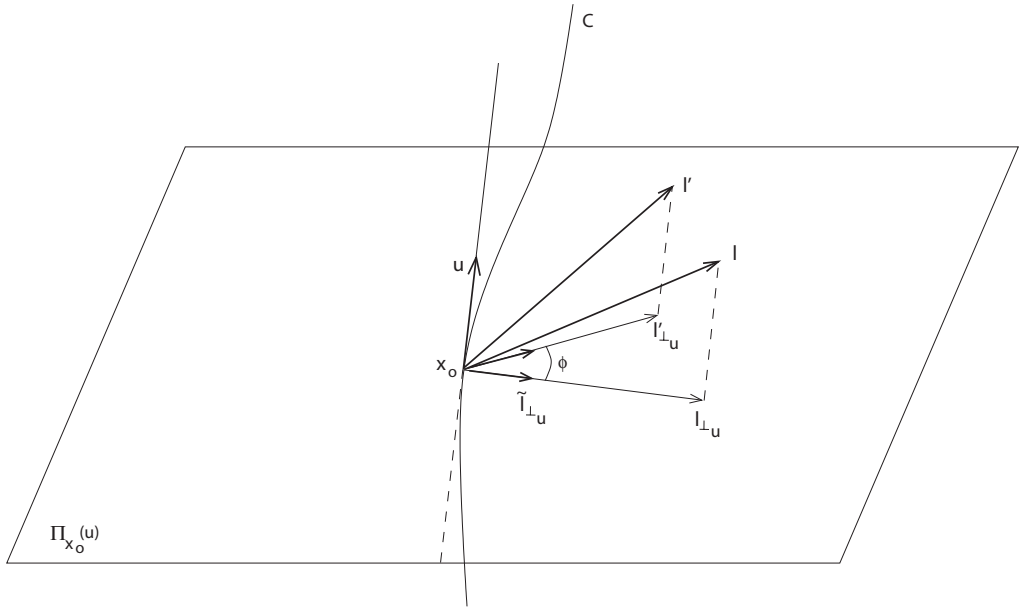}
      \caption{Angular distance as measured  by an observer at point $x_o$}
       \label{figure_mafig}
   \end{figure}
   
An arbitrary vector $V$ at $x_{o}$ admits one and only one decomposition of the form
\be \lb{pr0}
V = V_{\parallel_{u}} + V_{\bot_{u}} \, , 
\ee
where $V_{\parallel_{u}}$ is colinear to the unit vector $u$ and $V_{\bot_{u}}$ is a vector of the three-space $\Pi_{x_{o}}^{(3)}(u)$. Since $V_{\bot_{u}}$ and $u$ are orthogonal, one has 
\be \lb{pr1}
V_{\parallel_{u}} = (u . V) u
\ee
and 
\be \lb{pr2}
V_{\bot_{u}} = V - (u . V) u \, .
\ee
The vector $V_{\bot_{u}}$ is called the (orthogonal) projection of $V$ onto the three-space relative to the observer ${\cal O}(u)$. Its magnitude $\mid\!V_{\bot_{u}}\!\mid = \sqrt{- V_{\bot_{u}} . V_{\bot_{u}}}$ is given by
\be \lb{le1}
\mid\!V_{\bot_{u}}\!\mid  = \sqrt{(u . V)^2 - V^2} \, .
\ee
The direction of vector $V$ as seen by the observer ${\cal O}(u)$ is the direction of the unit spacelike vector $V_{\bot_{u}}^{\ast}$ defined as
\be \lb{di1}
V_{\bot_{u}}^{\ast} = \frac{V_{\bot_{u}}}{\mid\!V_{\bot_{u}}\!\mid} = \frac{V - (u . V) u }{\sqrt{(u . V)^2 - V^2}} \, .
\ee
Consider now a light ray $\Gamma$ received at $x_{o}$ and denote by $l$ a vector tangent to $\Gamma$ at $x_{o}$. In this work, we always assume that a vector tangent to a light ray is a null vector and is future oriented, so that
\be \lb{nul}
l^2 = 0 \, , \qquad \quad u.l >0 \, .
\ee
The direction of the ray $\Gamma$ as measured by the observer ${\cal O}(u)$ is the direction of the vector $l_{\bot_{u}}^{\ast}$. By using Eq. (\ref{di1}), and then taking into account Eqs. (\ref{nul}), it is easy to see that   
\be \lb{di2}
l_{\bot_{u}}^{\ast} = \frac{l}{u . l} - u \, .
\ee
Let $\Gamma'$ be an other light ray received at $x_{o}$. If $l'$ denotes a vector tangent to $\Gamma'$ at $x_{o}$, the direction of $\Gamma'$ as observed by ${\cal O}(u)$ is given by Eq. (\ref{di2}) in which $l'$ is substituted for $l$. As a consequence, the angular separation between $\Gamma$ and $\Gamma'$ as measured by ${\cal O}(u)$ may be defined as the angle $\phi_u$ between the two vectors $l_{\bot_{u}}^{\ast}$ and $l_{\bot_{u}}'^{\ast}$ belonging to the same subspace $\Pi_{x_{o}}^{(3)}(u)$ (see Soffel 1988, Brumberg 1991). Angle $\phi_u$ may be characterized without ambiguity by relations as follow
\be \lb{an1}
\cos \phi_u = - l_{\bot_{u}}^{\ast} . \, l_{\bot_{u}}'^{\ast} \, , \qquad 0 \leq \phi_u < \pi \, .
\ee
Taking into account properties of a light ray, we have the following relations
\begin{equation}
\label{llll}
l^2=g_{\mu\nu}l^\mu l^\nu\, , \quad l'^2=g_{\mu\nu}l'^\mu l'^\nu\, .
\end{equation}
As $u$ is an unitary vector, we can express $(1/u^0)^2$ with $g_{\mu\nu}u^\mu u ^\nu=1$ as follows
\be \lb{u0}
\frac{1}{(u^0)^2} = g_{00} + 2 g_{0i} \beta^{i} + g_{ij}\beta^{i}\beta^{j} \, ,
\ee
where
\be \lb{hlb}
\beta^{i} = \frac{dx^i}{dx^0} = \frac{dx^i}{c dt}  \, . 
\ee
Finally, substituting equations (\ref{llll}-\ref{hlb}) into equation (\ref{an1}) yields  the fundamental formula
\be \lb{an7}
\sin^2 \frac{\phi_u}{2} = - \frac{1}{4}\left[\frac{1}{(u^0)^2} \,  \frac{g^{ij}(\widehat{l}_{i}' - \widehat{l}_{i})(\widehat{l}_{j}' - \widehat{l}_{j})}{(1+ \beta^{m}\, \widehat{l}_{m})(1+\beta^{r}\,\widehat{l}_{r}')}\right]_{x_{o}} \, ,
\ee
where
\be \lb{ll3}
\widehat{l}_0 = 1\, , \,  \widehat{l}_i = \frac{l_i}{l_0}\, , \qquad  \widehat{l}_0' = 1\, , \,  \widehat{l}_i' = \frac{l_i'}{l_0'} \, .
\ee
The determination of the angular distance thus requires explicit computations of the ratios $\widehat{l}_i'$ and $\widehat{l}_i$. They can be obtained by the integration of null geodesic equations. However we will show that they result easily from the knowledge of the Synge's World function.
 
\subsection{World function and relativistic astrometry}

Synge's world function is a scalar function of the base point $x_o$ and the field point $x_s$. It is defined by (Synge, 1964)
\begin{equation}
\Omega(x_o,x_s)=\frac{1}{2}\int_0^1 g_{\mu\nu}(x^\alpha(\lambda))\frac{dx^\mu}{d\lambda}\frac{dx^\nu}{d\lambda}d\lambda\, ,
\end{equation}
and the integral is evaluated on the unique geodesic $\Gamma_{os}$ that links $x_o$ to $x_s$, $\lambda$ being an affine parameter. A fundamental property of $\Omega$ is to give an important information concerning the covariant components of the vectors tangent to $\Gamma_{os}$ at $x_o$ and $x_s$ respectively :
\begin{eqnarray}
\left(g_{\mu\nu}\frac{dx^\nu}{d\lambda}\right)_{x_o}&=&-\frac{\partial \Omega}{\partial x_o^\mu}(x_o,x_s)\, , \\
\left(g_{\mu\nu}\frac{dx^\nu}{d\lambda}\right)_{x_s}&=&\frac{\partial \Omega}{\partial x_s^\mu}(x_o,x_s)\, .\lb{rat2}
\end{eqnarray}
If $\Gamma_{os}$ is a light ray, we can consider $x_o$ as the observation point. Moreover, in this case, we have 
\begin{equation}
\Omega(x_o,x_s)=0\,.
\end{equation}
We recently show that explicit determination of $\Omega$ can be obtained from the integration of Hamilton-Jacobi equations without the knowkledge of $\Gamma_{os}$ (Le Poncin-Lafitte {\it et al.} 2004). All this means that determination of ratios $\widehat{l}_i'$ and $\widehat{l}_i$ require the following steps :
\begin{itemize}
\item{to determine $\Omega$ by solving Hamilton-Jacobi equations,}
\item{to impose the condition $\Omega=0$, }
\item{to compute the following relation 
\begin{equation}\lb{cov}
\widehat{l}_i=\frac{\partial \Omega/\partial x_o^i}{\partial \Omega/\partial x_o^0}\, .
\end{equation}}
\end{itemize}

\section{Conclusions}

In this paper, we give the general and rigorous expression of the observed angular distance between two point sources of light. The fundamental formulae (\ref{an7}) and (\ref{cov}) show that the theoretical calculation of the angular distance can be carried out when the world function is known. In this idealized sense, one can say that the problem of space astrometry involves one and only one unknown function. An other important point is that the aberration and the gravitational deflection of light cannot be treated as completely distinct phenomena. 


\end{document}